\documentclass[a4paper,11pt]{article}
\pdfoutput=1 

\usepackage{jcappub} 

\usepackage[T1]{fontenc} 

\newcommand{\mpl}{M_{\text{pl}}}

\newcommand{\bfx}{{\bf{x}}}

\newcommand{\gagg}{g_{a\gamma \gamma}}
\newcommand{\mns}{M_{\text{NS}}}

\title{\boldmath Resonant Conversion of Dark Matter Oscillons in Pulsar Magnetospheres}

\author[a,1]{Anirudh Prabhu,\note{Corresponding author.}}
\author[a]{and Nicholas M. Rapidis}

\affiliation[a]{Stanford Institute for Theoretical Physics,\\Stanford University, Stanford, California 94305, USA}

\emailAdd{aniprabhu@stanford.edu}
\emailAdd{rapidis@stanford.edu}

\arxivnumber{2005.03700}

\abstract{Due to their high magnetic fields and plasma densities, pulsars provide excellent laboratories for tests of beyond Standard Model (BSM) physics. When axions or axion-like particles (ALPs) approach closely enough to pulsars, they can be resonantly converted to photons, yielding dramatic electromagnetic signals. We discuss the possibility of detecting such signals from bound configurations of axions, colliding with pulsar magnetospheres. We find that all but the densest axion stars, \emph{oscillons}, are tidally destroyed well before resonant conversion can take place. Oscillons can be efficiently converted to photons, leading to bright, ephemeral radio flashes. Observation of the galactic bulge using existing (Very Large Array and LOFAR) and forthcoming (Square Kilometer Array) radio missions has the potential to detect such events for axion masses in the range $m_a \in \left[0.1 \ \mu\text{eV}, 30 \ \mu\text{eV}\right]$, even if oscillons make up a negligible fraction of dark matter.}

\begin{document}
\maketitle
\flushbottom

\section{Introduction}

Numerous astrophysical and cosmological observations suggest that a majority of the matter content of the universe is cold and non-baryonic. Understanding the non-gravitational properties of this dark matter (DM) remains one of the most important unsolved problems in physics. One of the best-motivated candidates for DM is the QCD axion. Originally introduced as part of a dynamical solution to the strong CP problem \cite{PhysRevLett.38.1440,PhysRevD.16.1791,PhysRevLett.40.223,PhysRevLett.40.279}, the QCD axion interacts weakly with the Standard Model (SM) and can be produced in the correct abundance to explain DM \cite{PRESKILL1983127,Abbott:1982af,Dine:1982ah}. Many beyond Standard Model (BSM) theories, and most notably string theory, also predict a plenitude of pseudoscalars, called axion-like particles (ALPs), which do not solve the strong CP problem but have similar properties and identical production mechanisms as the QCD axion and can also constitute the DM \cite{Arias:2012az}. In this paper, the term \textit{axion} will refer to general axion-like particles rather than just the QCD axion.

Depending on the production mechanism, axions can coalesce into dense structures, with densities much greater than the average local dark matter density. Such dense structures rely on the enhancement of axion perturbations relative to standard adiabatic perturbations. This can be realized by either $\mathcal{O}(1)$ fluctuations set by post-inflationary Peccei-Quinn (PQ) symmetry breaking, leading to \emph{axion miniclusters} \cite{Hogan1988,tkachev1986,KolbTkachev93,KolbTkachev941,KolbTkachev942,Vaquero:2018tib,Buschmann:2019icd,Eggemeier:2019khm} or by the parametric resonant growth of modes that occurs when the initial misalignment angle is large \cite{marios2019,Fukunaga:2020mvq}. In some situations, extremely dense objects called \emph{solitons} \cite{Kaup1968,Ruffini69, Tkachev1991, ChavanisI, BarrancoBernal, baum2017} and \emph{oscillons} \cite{DenseAxionStar, Braaten2017, baum2017} can form either through gravitational cooling of less dense objects or directly by self-interaction driven collapse of overdensities. Since parts of the discussion apply to general axion clumps, we use the term \emph{axion star} to refer to any such objects. When we intend to discuss a particular type of clump we will use the more specific classification. 

Most axion star models to date do not make robust, theoretical predictions about the fraction of dark matter that is bound up in axion stars, $f$. It is possible that $f \approx 1$, which would have serious implications for some direct detection experiments whose signal depends on the local dark matter density  \cite{PhysRevLett.124.101303,PhysRevD.97.092001,TheMADMAXWorkingGroup:2016hpc,PhysRevD.92.021101,Lee:2020cfj,Chaudhuri:2014dla,casper2017}. For most axion star models, detection could occur only during rare encounters between axion stars and the Earth.

The basis for many direct detection experiments is the axion-photon coupling.
Due to this coupling, axions can produce powerful electromagnetic signatures when they enter regions of strong magnetic fields. The conversion between axions and photons may be enhanced in a plasma, where the photon attains an effective mass. If the photon mass in the plasma is equal to the axion mass, the two particles will have the same dispersion relation, and resonant conversion can occur. Such conditions may be achieved in the magnetospheres of neutron stars. The effect of (resonant and non-resonant) axion-photon conversion in neutron star magnetospheres has been studied for homogeneous clouds of axions \cite{Anson2018,SSZ2019,Battye:2019aco,Leroy:2019ghm,Foster:2020pgt}, including the case of neutron star-black hole binary inspirals \cite{Edwards:2019tzf}, as well as for different types of axion stars \cite{Tkachev:2014dpa, Tkachev2015,iwazaki2014,PhysRevD.91.023008,iwazaki2017,Raby2016,Pshirkov2017,Bai2018, Dietrich2018}. Many of these results rely on the close approach of axion stars to the neutron star, which may not be possible due to tidal forces. In this work we circumvent these problems by considering the interactions of the densest axion stars in neutron star magnetospheres under the aforementioned resonant condition.\footnote{During the final stages of preparation of this manuscript, \cite{Buckley:2020fmh} was released. We discuss parts of their work in Section \ref{sec:FRB}.} In studying the conversion of these axion stars into photons, we consider the tidal forces that the neutron stars exert on the axion stars. Since most configurations of axion stars are tidally stripped in this scenario, we focus our attention on \textit{oscillons}, the densest axion stars, which are bound through self-interactions. While oscillons have a well-formulated formation method \cite{marios2019}, the longevity of such objects has not been firmly established. We discuss the lifetime of oscillons in Section \ref{sec:axionstarmodels}. We then consider the power produced by oscillon-neutron star (ONS) collisions as well as the rate at which these events occur in Sections \ref{sec:oscNSphys} and \ref{sec:rates} respectively. In Section \ref{sec:tidal}, we motivate the study of oscillons by discussing their tidal stability during these events. In Section \ref{sec:sensitivity}, using specifications from current radio missions, we estimate their sensitivity to ONS collisions, dependent on the fractional portion of the dark matter that they make up. While it has been postulated that fast radio bursts can be attributed to axion star-neutron star interactions \cite{Pshirkov2017,Raby2016,iwazaki2017,Iwazaki:2018squ, Buckley:2020fmh}, we explain why dense axion stars cannot be the source of all observed fast radio bursts. This is discussed in Section \ref{sec:FRB}.

 \section{Axion Star Models}
 \label{sec:axionstarmodels}
 
Axion stars and related objects form through the collapse of axion overdensities in the early universe. There exist many proposals for the formation of such overdensities. If PQ symmetry is broken after inflation, there will be $\mathcal{O}(1)$ fluctuations in the axion field on scales of order the horizon size at symmetry breaking. These fluctuations undergo gravitational collapse to form \emph{axion miniclusters} \cite{Hogan1988,tkachev1986,KolbTkachev93,KolbTkachev941,KolbTkachev942,Vaquero:2018tib,Buschmann:2019icd,Eggemeier:2019khm,Feix:2020txt}. Eventually the gravitational collapse is halted by kinetic pressure, leading to a dense central core, called a \emph{soliton}. In another scenario, when the axion is initially aligned close to the top of its potential, attractive self-interactions can lead to delay in the onset of oscillations of the axion field. In this situation, semi-relativistic modes can undergo parametric resonant growth, sourced by the axion zero-mode and curvature fluctuations. When these modes collapse, they form halos with considerable density enhancements over $\Lambda$CDM halos \cite{marios2019}. When the initial misalignment angle is tuned very close to the top of its potential\footnote{For periodic potentials, such as the cosine potential that arises from instanton effects, this means tuning the initial misalignment angle $\pi - \Phi_0 \ll 1$. For non-periodic potentials, it means $\Phi_0 \gg 1$. }, modes can go nonlinear deep into radiation domination, leading to self-interaction driven collapse into extremely dense objects called \emph{oscillons}. These objects are the focus of this paper.

Before discussing some generic properties of oscillons, we clarify the nomenclature used to describe them and related objects. We use the term ``soliton'' to describe objects in which the outward kinetic pressure is balanced by gravity. These are similar to \emph{bose stars}, introduced in \cite{Kaup1968, Ruffini69}. Bose stars were shown to collapse above a critical mass $M^*_\text{sol} \sim 1/(G m_a)$ \cite{BREIT1984329,UrenaLopez2002}. They also have a well-defined mass-radius relationship \cite{ChavanisI}

\begin{align}
    R_{99} \simeq \frac{10}{G M_a m_a^2}
\end{align}

\noindent where $R_{99}$ is the radius containing 99\% of the mass of the star. \emph{Axion stars} are bose stars made up of axions. At masses much lower than the critical mass, the properties of axion stars and bose stars coincide. Near the critical mass, however, the axion potential can induce weak, attractive self-interactions that can alter the balance of forces. Axion stars are described by solutions to the Gross-Pitaevskii-Poisson (GPP) equation, with critical axion star mass given by \cite{ChavanisI}

\begin{align}
    M^*_{\text{as}} &= \frac{10.15}{\sqrt{|\lambda_4|}} \frac{\mpl f_a}{m_a} \label{eqn:mcritdas}
\end{align}

\noindent where $\lambda_4$ is an $\mathcal{O}(1)$ dimensionless coupling constant associated with the quartic term in the potential. For the QCD axion with a chiral potential, $\lambda_4 = -0.34$. We consider models with attractive self-interactions for which $\lambda_4 < 0$. Axion stars with masses less than (\ref{eqn:mcritdas}) are known as \emph{dilute axion stars}. Above the critical mass (\ref{eqn:mcritdas}), an axion star can collapse to form either sub-critical axion stars or black holes, or simply radiate away relativistic axions \cite{LevkovCollapse, Helfer_2017, Widdicombe_2018}. It has also been proposed that near-critical axion stars may be unstable to photon perturbations, due to parametric resonant growth of photon modes in an axion background \cite{Hertzberg_2018,levkov2020radioemission}, however this usually requires enhanced axion-photon couplings $\gagg \gtrsim \mathcal{O}(0.1)/f_a$. 

When the axion amplitude becomes $a(x)/f_a \sim 1$, the quartic term alone no longer suffices to describe self-interactions. The resultant objects are meta-stable solutions to the Sine-Gordon equation called \emph{oscillons}. When the constituent bosons are axions, they are known as \emph{dense axion stars} \cite{DenseAxionStar, baum2017}. Currently, these objects are understood to have very long (compared to the axion oscillation period), but nevertheless finite lifetimes \cite{PhysRevD.49.2978,Salmi:2012ta,baum2017,Fodor:2006zs,Bogolyubskii1976}.

Crucial to the observation of oscillons is their present day abundance, which is determined by their production mechanism and lifetime. Understanding the oscillon lifetime remains an open problem. Oscillons comprised of axions with cosine potentials have been shown to decay after $\tau \sim 10^3 \  m_a^{-1}$ \cite{baum2017,OscillonDarkMatter}. For our purposes, this will lead to oscillon decay well before matter-radiation equality. The issue of lifetime can be ameliorated by considering models with flat potentials. Flat potentials can be generated naturally in theories such as those described in \cite{eva2011,Geng:2019phi}. Consider a general class of potentials \cite{Dubovsky2011}

\begin{align}
    V(a) &= \frac{m^2 f_a^2}{2 p} \left[ \left(1 + \frac{a^2}{f_a^2} \right)^p - 1 \right]. \label{eqn:potential}
\end{align}
Here, models with $0 \le p \le 1 $ give axion-monodromy-like potentials \cite{Eva2008, EvaMcAllister2008} and $p<0$ give potentials that are flat at large field values. Oscillons in the former category were shown to decay after a time $\tau \sim 10^8 m_a^{-1}-10^{9} m_a^{-1}$, while for $p<0$, oscillons were shown to persist for time $\tau \gtrsim 10^{9} m_a^{-1}$ with no sign of decay \cite{OscillonDarkMatter,Zhang:2020bec}. This serves as a lower bound, leaving open the possibility of oscillons being long-lived even on cosmological timescales. Another possibility that warrants investigation is that a significant oscillon population is formed through the recent mergers and subsequent collapses of stable, near-critical solitons. Motivated by these possibilities, we assume a present-day abundance of oscillons and investigate the observational signatures of their collisions with neutron stars. We quote some general characteristics of oscillons in the context of dense axion stars \cite{baum2017} which have also arisen over a wide variety of initial conditions in simulations performed in \cite{marios2019}. The first is that in the oscillon branch, the central energy density is set by the natural scale, $\rho_0 \sim m_a^2 f_a^2$. They were also seen to have size $R_\text{osc} \sim n/m_a$ where $n \lesssim \mathcal{O}(10)$ and approximately flat internal density profile. This gives an oscillon mass $M_{\text{osc}} \sim \rho_0 R^3 = n^3 {f_a}^2/m_a$. Finally, the spectrum of these objects has peaks at $\omega_\text{peak} \approx (2n+1)\omega_0, \ n=0,1, 2, \dots$, where $\omega_0$ is the fundamental frequency of the oscillon. This spectral feature highlights the oscillon metastability, which is thought to be the result of number-changing processes.

\section{Axion-photon mixing \label{section:axion-photon}}

Interactions between the axion field, $a(x)$ and electromagnetic fields, $F_{\mu \nu}$ are described by the interaction Lagrangian

\begin{align}
\mathcal{L} \supset -\frac{1}{4}g_{a\gamma\gamma}a F_{\mu\nu}\tilde{F}^{\mu\nu}=g_{a\gamma\gamma}a\mathbf{E}\cdot \mathbf{B} 
\end{align}

\noindent where $\tilde{F}^{\mu\nu} = \epsilon^{\mu \nu \alpha \beta} F_{\alpha \beta}/2$ and $\gagg$ is the model-dependent axion-photon coupling. For the QCD axion, it is given by 

\begin{align}
    \gagg = \frac{\alpha}{2\pi f_a} \left( \frac{E}{N} - 1.92 \right)
\end{align}

\noindent where $E$ and $N$ are the electromagnetic and color anomalies associated with the axion axial current, respectively. In the DFSZ model, $E/N = 8/3$ \cite{DFOG,ZOG} and in the KSVZ model, $E/N = 0$ \cite{KimOG,SVZOG}, but these quantities can take on a range of $\mathcal{O}(1)$ values. There exist exotic models in which the electromagnetic anomaly can be exponentially enhanced, leading to a large axion-photon coupling \cite{photophilic2016}. Calculating the axion-photon coupling for our models of interest is beyond the scope of this paper and we assume $\gagg = \alpha/(2\pi f_a)$.  

The axion-photon interaction is widely used in experimental searches for axions, where axions are expected to interact with an external magnetic field to produce detectable photons \cite{PhysRevLett.124.101303,PhysRevD.97.092001,TheMADMAXWorkingGroup:2016hpc,PhysRevD.92.021101,Lee:2020cfj,Chaudhuri:2014dla}. Since the signal power of an axion-to-photon conversion scales quadratically with the magnetic field, it is preferable to maximize the magnetic field. To detect such axions, one can observe astrophysical objects which can sustain large magnetic fields. Some of the largest magnetic fields in nature arise in magnetospheres of rotating neutron stars, with some surface fields reaching nearly $B = 10^{15}$ G. The rotating magnetic field induces a strong electric field at the pulsar surface, which drags charged particles along the magnetic field lines. This results in a plasma magnetosphere that co-rotates with the pulsar \cite{GJ1969}. In pulsar magnetospheres, the photon develops an effective plasma mass $\omega_p = \sqrt{4\pi \alpha n_e /m_e}$ where $n_e$ is the electron density of the plasma. When axions fall into the pulsar they will be resonantly converted to photons at a distance where the effective photon mass coincides with the axion mass \cite{Anson2018,Battye:2019aco,Leroy:2019ghm,Foster:2020pgt}. To understand this resonance, we consider the axion-photon mixing equations in the presence of a magnetic field and a background plasma \cite{Raffelt:1987im}.

Around the conversion radius, oscillon-neutron star (ONS) collisions can be understood using a simplified model of an axion plane-wave converting to an electromagnetic plane-wave propagating in the same direction. In reality, axions within the oscillon have a moderately relativistic velocity dispersion, $\sigma_v$, which is  comparable to or larger than the velocity of the oscillon when it crosses the conversion region, $v_c$. Thus near the conversion radius, oscillons can be decomposed into plane waves propagating in all directions. Assuming a constant external magnetic field, the various modes decouple from each other, leading to an approximately isotropic signal. Near the conversion radius, the axion and photon fields may be written in the form  $A_\parallel(z,t)=e^{i(\omega t- kz)}A_{\parallel,0}(z)$ and $a(z,t)=ie^{i(\omega t- kz)}a_0(z)$ where $z$ is the direction of propagation, $A_\parallel$ is the component of the photon field parallel to the magnetic field and $k^2 = \omega^2 - m_a^2$. The mixing equation amongst the component of the photon perpendicular to the magnetic field $A_{\bot,0}$, the component parallel to magnetic field $A_{\parallel,0}$, and the axion, $a_0$ is \cite{Raffelt:1987im}

\begin{equation}
\left[\omega^2+\partial_z^2+\begin{pmatrix}Q_\bot & 0 & 0 \\ 0 & Q_\parallel& B_T g_{a\gamma\gamma} \omega\\ 0 & {B_Tg_{a\gamma\gamma}\omega} & -m_a^2\end{pmatrix}\right]\begin{pmatrix} A_{\bot,0}\\ A_{\parallel,0} \\ a_0
\end{pmatrix}=0 \label{eqn:3dmix}
\end{equation}
where $B_T$ is the transverse magnetic field, $\omega$ is the photon frequency, $m_a$ is the axion mass, and $Q_\bot$ and $Q_\parallel$ depend on the physics of the medium in which the photon is propagating. Working in the WKB approximation where the spatial gradient of the magnetic field is smaller than the wavelength of the photon and the axion, we write $\omega^2+\partial_z^2=(\omega+i\partial_z)(\omega-i\partial_z)=(\omega+k)(\omega-i\partial_z).$ Ignoring the perpendicular component of the photon which does not mix with the axion and working in the case of a medium with a plasma mass, such as that of the Goldreich-Julian (GJ) neutron star, the mixing equations become

\begin{equation}
\left[-i\partial_z+\begin{pmatrix} \Delta_\parallel& \Delta_M\\  \Delta_M & 0\end{pmatrix} \right]\begin{pmatrix}A_{\parallel,0}(z)\\ a_0(z)\end{pmatrix}=0
\label{eq:mixing_matrix}
\end{equation}

\noindent with

\begin{equation}
\Delta_{\parallel}=\frac{1}{2k}\left[m_a^2-\omega_p^2\frac{\sin^2\theta}{1-\left(\frac{\omega_p\cos \theta}{\omega}\right)^2}\right],\quad \Delta_M=\frac{B_Tg_{a\gamma\gamma}m_a}{2k}\frac{\sin\theta}{1-\left(\frac{\omega_p\cos \theta}{\omega}\right)^2}
\end{equation}
where $\omega_p$ is the plasma mass of the photon and $\theta$ is the angle between the magnetic field and the propagation direction. Here $\omega = m_a\sqrt{1+v_c^2}$ where $v_c \simeq \sqrt{2 G \mns/r_c}$ is the oscillon velocity at the conversion radius. Since the axion-photon mixing depends on the transverse magnetic field, any virialization of the axions within the oscillon will lead to an $\mathcal{O}(1)$ suppression of the signal due to the relative angle between the propagation direction of the axion and the magnetic field.

The energy transfer function can be extracted from Equation \ref{eq:mixing_matrix}. For a radially infalling axion, where $B_T(r)=B(r)$ at a radius $r$, the ratio of the photon energy density to the axion energy density is \cite{Anson2018}

\begin{equation}
p_{a\gamma}(r)=\frac{|A_\parallel|^2}{|a_0|^2}=\left|\int_0^{r}dr'\frac{g_{a\gamma\gamma}B(r')\xi(r')}{2v_c\sin\tilde{\theta}}\exp\left[\frac{-i\int_0^{r'}d\tilde{r}(m_a^2-\xi(\tilde{r})\omega_p^2(\tilde{r}))}{2m_av_c}\right]\right|^2.\\
\end{equation}
Upon evaluating the above expression, the energy transfer function is given by
\begin{equation}
p_{a\gamma}^\infty=v_c\lim_{r\to \infty}p_{a\gamma}(r)=\frac{\pi r_c g_{a\gamma\gamma}^2B(r_c)^2}{3m_a}.
\end{equation}
where $r_c$ is the conversion radius of the neutron star. Considering the magnetosphere at an angle $\pi/2$ with respect to the NS rotation axis, we define a radial distance $L$ over which the resonant conversion happens:
\begin{equation}
    L=\sqrt{\frac{2\pi r_c v_c}{3m_a}}. \label{eqn:conversionthickness}
\end{equation}
Since the change in the resonant frequency is negligible in this region, the total energy emitted by an ONS collision can be obtained by integrating the power over the time it takes to cross the region $L$. The physics of the conversion radius is discussed in the following section. 

The discussion above neglects the non-resonant conversion that takes place before the axion star reaches the conversion radius, and after it has passed through. At a distance $r>r_c$, the conversion probability is less than that at the conversion radius by an amount \cite{SSZ2019}

\begin{align}
    \frac{p^{\text{non-resonant}}_{a\gamma}(r)}{p^{\text{resonant}}_{a\gamma} } \sim \left(\frac{r_c}{r} \right)^6 {\frac{1}{m_a r_c}} \ll 1.
\end{align}
 
In the region $r_{\text{NS}}<r<r_c$, the photon mass becomes greater than the axion energy, which kinematically suppresses axion-photon conversion. Based on the arguments above, we conclude that radio signals from ONS collisions occur predominantly in the resonant regime.

\section{Oscillon Neutron Star physics \label{sec:oscNSphys}}

Neutron star properties are largely inferred from those of the $\sim 2000$ pulsars observed and catalogued by the Australia Telescope National Facility (ATNF) \cite{ATNF}. These \emph{active} pulsars are characterized by their extremely bright, pulsed emission, making them good candidates for observation. However, there exists a much larger population of \emph{dead} neutron stars that do not emit observable radiation. Population models of dead NSs depend sensitively on the evolution of spin periods and magnetic fields, which have large uncertainties \cite{SSZ2019}. We optimistically adopt the model considered in \cite{Model1} which, in contrast to models such as \cite{Model2}, assumes magnetic fields do not decay appreciably over the lifetime of the NSs. In \cite{Model1} NS magnetic fields, measured in Gauss, are log-normally distributed with (population) average surface magnetic field $\left\langle B_0 \right\rangle \approx 10^{13}$ G and initial period $\left\langle P \right \rangle \approx 0.3$ sec. The rotation period of all NSs decrease over time due largely to dipole radiation. Simulations in \cite{SSZ2019} predict a present-day population of NSs with average period $\left\langle P \right \rangle \approx 1$ sec. For simplicity, we assume all present-day NSs to have the average period and magnetic field mentioned above. A thorough consideration of the distribution of magnetic fields and periods will yield $\mathcal{O}$(1) changes to our results. We emphasize that if, as in \cite{Model2}, magnetic field decay occurs, the results can change drastically. We do not consider this possibility in this paper.

In the GJ model\footnote{As discussed in \cite{SSZ2019}, dead NSs may be better described by the electrosphere model, which differs qualitatively from the GJ model. Compared to the GJ model, the electrosphere model predicts a more rapid decay of the plasma density far from the NS surface. Considering this model may reduce sensitivity to low axion masses.}, the pulsar magnetic field is modeled as a magnetic dipole that co-rotates with the star. Due to the strong induced electric field, charged particles are forced to co-rotate with the pulsar as well. The non-zero charge density in the pulsar magnetosphere gives the photon an effective mass

\begin{equation}
    \omega_p\approx 150 \text{ GHz} \left(\frac{B_z}{10^{14}\text{ Gauss}}\right)^{1/2}\left(\frac{P}{1\text{ sec}}\right)^{-1/2}.
\end{equation}

\noindent where $B_z$ is the $z$-component of the magnetic field and $P$ is the pulsar period. Utilizing the results from Section \ref{section:axion-photon} it is possible to extract a parametric form for the location of the conversion radius in the neutron star magnetosphere

\begin{equation}
\begin{split}
    r_c &= 169 \text{ km} \left|3\cos\theta \cos(\theta_m-\theta)-\cos(\theta_m)\right|^{1/3}\left(\frac{r_0}{10\text{ km}}\right)\\
    &\hspace{1.3in} \times \left(\frac{B_0}{10^{14}\text{ G}}\right)^{1/3}
    \left(\frac{P}{1\text{ sec}}\right)^{-1/3}\left(\frac{m_a}{10^{-6}\text{ eV}}\right)^{-2/3}
\end{split}
\end{equation}

\noindent where $\theta_m$ is the angle between the magnetic dipole and the orbital axis of the pulsar. This is the radial location at which the maximum power will be emitted from the oscillon as axions convert into photons. The resonant conversion will only occur, however, if the conversion radius is within the pulsar light cylinder, which has radius $R_{\text{LC}} = P/2\pi$ where $P$ is the pulsar period. This is only an issue if we consider axions on the low end of the mass window colliding with millisecond pulsars with extremely high magnetic fields. While there will be non-resonant conversion of axions to photons in the region between the edge of the light cylinder and the conversion radius, the fractional loss of the oscillon mass within this window is $\lesssim 10^{-14}$ for the oscillon parameter space of interest.

For axions of mass $10^{-7}\text{ eV}\lesssim m_a\lesssim 10^{-3}\text{ eV},$ the radius of the oscillon is much smaller than that of the neutron star, $ r_{\text{NS}}$. This means that any curvature in the magnetic field is negligible over a length scale of $R_{\text{osc}}$. Hence the magnitude of the magnetic field is constant and is given by
\begin{equation}
B(r_c)=B_0\left(\frac{r_0}{r_c}\right)^3\sqrt{3\cos^2(\theta-\theta_m)+1}
\end{equation}
where $\theta$ is an arbitrary angle relative the the axis of rotation.

The maximum power of the interaction is produced when the center of the oscillon is situated at $r_c$. Averaged over all such angular positions on the neutron star, the power of the event is given by the flux times the power transfer function
\begin{equation}
\begin{split}
P&=p_{a\gamma}^\infty\rho_{\text{osc}}v_c\pi R_{\text{osc}}^2\\
&= 6.7\times 10^{27}\text{ W}  \left(\frac{B_0}{10^{14}\text{ G}}\right)^{1/6}\left(\frac{r_0}{10\text{ km}}\right)^{1/2}\left(\frac{M_{\text{NS}}}{1.4\ M_\odot}\right)^{1/2}\left(\frac{n}{10}\right)^2\left(\frac{m_a}{10^{-6}\text{ eV}}\right)^{8/3}
\end{split} \label{eqn:power}
\end{equation}
where $n$ is the numerical coefficient determining the radius of the oscillon, $R_\text{osc} \sim n /m_a$, and $g_{a\gamma\gamma}=\alpha/2\pi f_a$. The velocity of the oscillon is given by $v_c=\sqrt{2GM_{\text{NS}}/r_c}$, implying that it falls radially and has no azimuthal component in the velocity. Furthermore, modifications to the shape and density of the oscillon due to tidal effects have been ignored here. Tidal effects are discussed in Section \ref{sec:tidal}.

The duration of these events is determined by $L/v_c$, where $L$ is the size of the conversion region as given in Equation \ref{eqn:conversionthickness}. For a neutron star of mass $1.4\ M_{\odot}$, radius 10 km, and magnetic field $10^{14}$ Gauss, the duration of the event, $\tau$, scales as
\begin{equation}
    \tau \sim 2\times 10^{-6}\text{ s}\times \left(\frac{m_a}{10^{-6}\text{ eV}}\right)^{-1}
    \label{eqn:eventduration}
\end{equation}
where it is assumed that the oscillon is falling radially inwards and has no azimuthal component to its velocity.

Depending on the values of $m_a$ and $f_a,$ all of the axions in an oscillon may be converted into photons as they traverse the resonant region of the magnetosphere. The fractional mass lost in one such encounter with a neutron star of mass 1.4 $M_\odot$ and radius 10 km scales as $10^{-7}(m_a/ 10^{-6}\text{ eV})^{8/3}(f_a/10^{10}\text{ GeV})^{-2}$. While some oscillons will be entirely converted into photons, this process does not deplete the DM halos since the rate of such events is sufficiently low and the DM halos are sufficiently massive.

The power from these events is, in principle, strong enough so that observations could be made using terrestrial radio telescopes. The flux density that will be observed on Earth from an ONS encounter at a distance $d$ from the Earth is given by

\begin{equation}\begin{split}
\Phi&=\frac{P}{4\pi d^2 (m_a/2\pi)}\\
& =8.2\times10^{4}\text{ Jy}\left(\frac{B_0}{10^{14}\text{ G}}\right)^{1/6}\left(\frac{r_0}{10\text{ km}}\right)^{1/2} \\
&\qquad \times \left(\frac{M_{\text{NS}}}{1.4\ M_\odot}\right)^{1/2}  \left(\frac{m_a}{10^{-6}\text{ eV}}\right)^{5/3}\left(\frac{d}{1\text{ kpc}}\right)^{-2} \left(\frac{n}{10}\right)^2
\end{split} \label{eqn:flux}
\end{equation}
where we have set the bandwidth of this signal to be $\Delta \omega \sim m_a/(2\pi)$ (a more detailed explanation of this assumption is given in Sec. \ref{sec:sensitivity}).  It is noteworthy that any $f_a$ dependence, and hence $g_{a\gamma\gamma}$ dependence, is absent from the expression. This is an artifact of fixing the axion photon coupling to be $\gagg = \alpha/(2\pi f_a)$.

\section{Oscillon-Neutron Star Encounter Rate \label{sec:rates}}

The fluxes associated with ONS collisions (\ref{eqn:flux}) are sufficiently large that we may expect to see events both in the Milky Way and nearby satellite galaxies. The potentially observable ONS encounter rate is given by 

\begin{equation}
\Gamma = \displaystyle\int_{V_{\text{obs}}} n_\text{osc}(\bfx) n_{\text{NS}} (\bfx)  \langle \sigma v\rangle d^3 \bfx 
\end{equation}

\noindent where $V_{\text{obs}}$ is the observing volume of the telescope, that is the total volume within the field-of-view (FoV), $n_{\text{osc}}(\bfx)$ and $n_{\text{NS}}(\bfx)$ are the local number densities of the oscillons and neutron stars respectively, and $v$ is the virialized velocity of the dark matter in the galaxy. The small FoV of most telescopes introduces a trade-off between near and far sources. While near sources (i.e. Galactic center) have higher observed fluxes, there will be fewer events within the FoV. Given this limited FoV, observations should be made as close to the center of the halo as possible. 

As mentioned above, for a fixed event luminosity, the optimal observational scheme is determined by the distance to the source, $D_S$ and the FoV of the observing radio telescope, $F$. Choosing a source will then depend crucially on the distribution of DM and neutron stars within a galaxy. The former has been well-modeled in $N$-body simulations, giving a density profile that is well-fitted by the Navarro-Frenk-White (NFW) profile over a wide range of halo masses \cite{NFW1997},

\begin{align}
\rho_{\text{DM}}(r) &= \frac{4 \rho_s}{\frac{r}{R_s}\left(1+\frac{r}{R_s}\right)^2}.
\end{align}

\noindent Here $\rho_s$ is the scale density and $R_s$ is the scale radius. Neutron stars are formed from core-collapse of massive stars or, on rare occasion, by the accretion of matter by White Dwarfs (WDs) that are very close to the Chandrasekhar limit. The number of NS in a halo can be approximated by the $N_{\text{NS}} = \gamma_{_C} \tau$ where $\gamma_{_C}$ is the present-day core-collapse rate and $\tau \sim 10$ Gyr is the age of the universe \cite{Diehl2006}. The core-collapse rate in the Milky Way is $\gamma_{_\text{MW}} = 2.84 \pm 0.6$ per century (with an overall factor of $\sim 2$ systematic uncertainty), leading to a population of $N_{_\text{MW}} \sim 10^9$ NS \cite{SNRate}. For the low-redshift satellite galaxies of interest, we assume similar core-collapse rates, and thus a number of NS that scales with halo volume. In the MW, a majority of NS reside in the bulge and disk with approximately 60\% being in the former and 40\% in the latter \cite{Ofek2009, Sartore2010}. The bulge proves to be the best candidate for observation due to its higher projected number density (in heliocentric coordinates) \cite{Sartore2010}. \footnote{We have neglected the nuclear star cluster, which could have a comparable or greater NS population \cite{Foster:2020pgt}.} We assume the neutron star number density in the bulge is given by 

\begin{align}
    n_{\text{bulge}}(r) &= \frac{N_{\text{bulge}}}{2\pi r^3} \frac{a}{r}\left(1 + \frac{a}{r} \right)^{-3}
\end{align}

\noindent where $N_\text{bulge} = 6 \times 10^8$ is the number of neutron stars in the bulge, $a \approx 0.6$ kpc is a scale parameter, and $r$ is the distance from the galactic center \cite{SSZ2019}. The cross section of these encounters, $\sigma$, will be enhanced due to the gravitational attraction of the two objects (this is often times referred to as Sommerfeld enhancement) so that the maximum impact parameter at which the oscillon traverses the conversion radius of the neutron star is 
$h=\sqrt{{2GM_{\text{NS}}r_c}}/{v_{0}}$. Since $h\gg r_\text{osc}$ in the $m_a$ regime of interest, $h$ determines the cross section of the encounter. A detailed study of the dynamics of axion miniclusters inside the Milky Way can be found in \cite{Dokuchaev:2017psd}. The expected event rate in the bulge is

\begin{align}
    \Gamma_{\text{yr}} &\approx 200 {\text{ yr}}^{-1}\left( \frac{m_a}{10^{-6} \text{ eV}} \right)^{1/3} \left( \frac{f_a}{ 10^{12} \text{ GeV}} \right)^{-2} \left(\frac{B_0}{ 10^{14} \text{ Gauss}} \right)^{1/3} \left( \frac{N_\text{bulge}}{6\times 10^8} \right)\nonumber \\
    &\qquad \times\left( \frac{P}{1 \text{ sec}} \right)^{-1/3} \left( \frac{\rho_s}{0.06 \text{ GeV}/\text{cm}^3} \right)^{1/3} \left(\frac{M_{\text{NS}}}{1.4 \ M_\odot}\right)
\end{align}

\section{Tidal Disruption} \label{sec:tidal}

In order for an axion star to resonantly convert to radiation, it must reach the critical radius without being tidally destroyed by the neutron star.\footnote{It is possible that even if the star is tidally disrupted, it will form a stream that will enter orbit around the neutron star and give rise to a signal with a lower peak flux but longer duration. This scenario will be studied in future work.} Previous claims that gravitationally bound solitons could be responsible for fast radio bursts \cite{iwazaki2014,PhysRevD.91.023008, Tkachev2015} do not fully consider tidal effects on these objects. Tidal stretching is considered in \cite{Raby2016, Pshirkov2017} but only for the case of solitons. In this section, we present a heuristic derivation of the tidal disruption radius for oscillons, showing that, for the mass range of interest, oscillons are not tidally destroyed before being converted. We also outline a more detailed calculation, which could be applied to solitons and less dense objects.

To study tidal disruption of axion stars, and in particular oscillons, we introduce the Roche radius, $r_{\text{roche}}$. For an object, with mass $M$, the Roche radius is defined as the maximal radial position from the mass $M$ at which its tidal forces dissociate a less massive object of mass $m$, whose radius is $r$. In particular, the Roche radius $d,$ solves $F_{\text{Tidal}}=F_{\text{bind}}$ where $F_{\text{bind}}$ is the binding force and

\begin{equation}
F_{\text{Tidal}}=G_NMm\left(\frac{1}{(d-r/2)^2}-\frac{1}{(d+r/2)^2}\right).
\label{eq:roche}
\end{equation}

For the case of an oscillon, the binding force is dominated by the self interactions of the axion rather than its self gravity. We assume the potential in Equation \ref{eqn:potential} and set $p=-1/2$ so that we have long-lived oscillons. This method, however, can be generalized for any arbitrary periodic potential. Given the potential, the self-interaction force can be obtained by taking the derivative of the potential energy of the system $U$

\begin{equation}
F_{\text{Self}}=-\nabla U.
\label{eq:selfforce}
\end{equation}

\noindent The potential energy $U$, contains a term from its self gravity, a kinetic pressure term, and a term from the potential $V(a)$. To integrate the kinetic pressure and potential over the volume of the oscillon we set the profile of the axion field within the oscillon, $a(r)$, to be a Gaussian of width $r_{\text{osc}}\approx n/(2.5m)$ and with a central value that is $\mathcal{O}(1)$. Then integrating and cutting off the field at $r=r_{\text{osc}}$ the potential energy has the form:

\begin{equation}
U=-\frac{GM_{\text{osc}^2}}{r_{\text{osc}}}+\int_{\text{osc}}dV\left[\frac{f_a^2}{2}\left(\frac{da(r)}{dr}\right)^2+V\big(a(r)\big)\right].
\end{equation}

\noindent Per Equations \ref{eq:roche}, \ref{eq:selfforce}, the Roche radius is the value $d$ which solves the following equation for different values of $m_a$ and $f_a$, 
\begin{equation}-\nabla U=GM_{\text{osc}}M_{\text{NS}}\left(\frac{1}{(d-r/2)^2}-\frac{1}{(d+r/2)^2}\right).
\end{equation}
Since $M_{\text{osc}}\propto f_a^2$ and the gravitational term in $U$ is heavily suppressed by the factor of $G$, the $f_a$ dependence is effectively removed from the system. Ultimately, for axion masses $10^{-7}\lesssim m_a \lesssim 10^{-3}\text{ eV}$, the conversion radius exists at distances far outside the Roche radius of the system at $r_c \gtrsim 10^5 r_{\text{roche}}$ and hence oscillons can reach the conversion radius intact. This also implies that any deformation to the profile of the oscillon due to tidal effects is negligible and can be ignored at the conversion radius.

\begin{figure}
    \centering
    \includegraphics[scale=0.6]{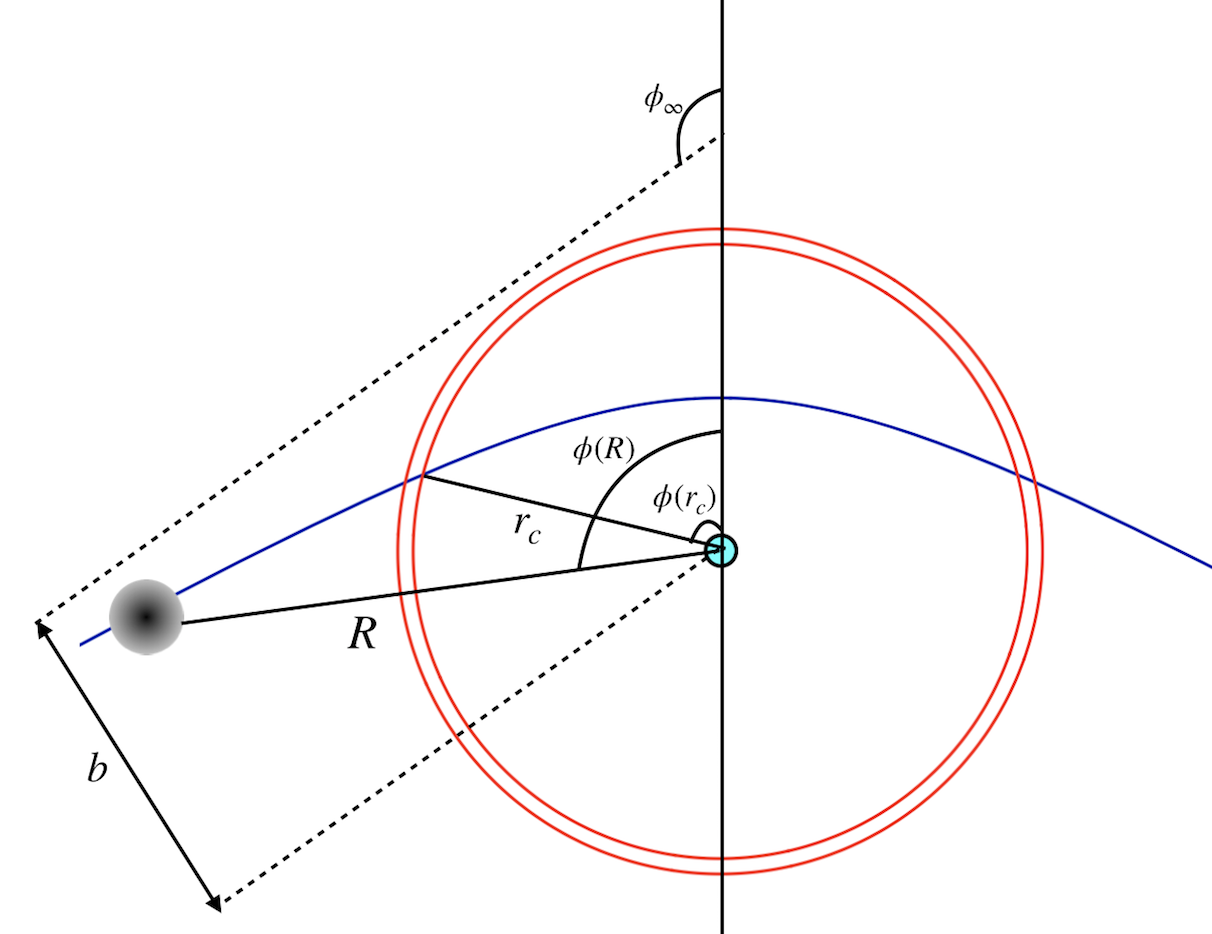}
    \caption{Graphical representation of the orbital dynamics of an axion star around a pulsar.}
    \label{fig:orbit}
\end{figure}

This heuristic calculation supports the idea that oscillons are stable against tidal disruption until they reach the conversion radius. A more detailed consideration involves calculating the differential velocity kick imparted by the neutron star during the trajectory of the oscillon as shown in Fig. \ref{fig:orbit}. When this differential velocity becomes comparable to the virial velocity of axions in the axion star, we say the star becomes tidally disrupted. In the impulse approximation, the differential velocity kick is given by

\begin{align}
    \Delta v_{\text{impulse}}(R) = 2 G \mns R_a \displaystyle\int_{\phi_\infty}^{\phi(R)} \frac{ r'(\phi) \hat{r}(\phi) }{ r(\phi)^3 } d\phi 
\end{align}

\noindent where $r(\phi)$ is the trajectory of the axion star in the orbital plane, $\hat{r}(\phi)$ is the unit vector pointing from the axion star to the neutron star, $\mns$ is the neutron star mass, $M_a$ is the axion star mass, $R_a$ is the axion star radius, $b$ is the impact parameter, $\phi_\infty$ is the true anomaly, and $\phi(R)$ is the polar angle of the trajectory when the axion star is at a distance $R$ from the neutron star. The impulse approximation, however, applies only when the duration of the gravitational shock, due to the neutron star, is small compared to the orbital period of axions in the object. When this condition is violated, the effect of the gravitational perturbation is smeared over many orbital periods, suppressing its effect. This can be corrected with the so-called ``adiabatic correction'' (also called the ``Weinberg correction''), $A(x)$ \cite{OstrikerCorrection}, which is well approximated by 

\begin{align}
    A(x) &= (1+x^2)^{-3/2}
\end{align}

\noindent where $x = \tau/T$. Here $\tau$ is the timescale over which the gravitational shock occurs and $T$ is the orbital period of axions in the axion star. Since, in oscillons, axions move at moderately relativistic velocities, the adiabatic correction introduces an enormous suppression to the tidal force of the neutron star.  It is worth noting that in some near-critical solitons, axion velocities may be almost relativistic, which could lead to an appreciable adiabatic suppression, compared to na\"{i}ve estimates from the impulse approximation. We leave a detailed study of tidal effects on solitons and less dense halos to future work.

 \section{Sensitivity Analysis \label{sec:sensitivity}}
 

The flux density of an ONS collision was calculated in (\ref{eqn:flux}). For a given telescope, the minimum detectable flux density is

\begin{align}
    S_{\text{min}} &= \text{SNR} \frac{\text{SEFD} }{ \sqrt{2 \mathcal{B} T}} \label{eqn:smin}
\end{align}

\noindent where SNR is the required signal-to-noise ratio, SEFD is the \emph{system equivalent flux density}, $\mathcal{B}$ is the observing bandwidth, and $T$ is the integration time. The factor of two comes from the number of polarizations observed. The radio data will be in the form of power measurements in $N_\omega$ frequency bins and $N_t$ time bins. Under the null hypothesis, we assume an expected flux density per bin of $S_0$, which is independent of frequency and time. In the presence of an axion signal, this expected power will be supplemented by a flux density $S_{\text{axion}}(i,j)$, which depends upon the signal spectrum and time curve.  Assuming the power measurements are independent and normally distributed, we can construct a $\chi^2$ test statistic,

\begin{align}
    \chi^2 = \displaystyle\sum_{i = 1}^{N_t} \displaystyle\sum_{j = 1}^{N_\omega} \frac{S_{\text{axion}}(i,j)^2}{ \sigma^2_S}.
\end{align}

\noindent where $\sigma^2_S$ is the flux density error in each bin, given by (\ref{eqn:smin}) with the bandwidth and integration time being the widths of each frequency and time bin, respectively. The resulting $\chi^2$ value is well-approximated by 

\begin{align}
    \chi^2 = \frac{2} { \text{SEFD}^2} \displaystyle\int_{\omega_{\text{min}}}^{\omega_{\text{min}} + \mathcal{B}} \displaystyle\int_{t_{\text{min}}}^{t_{\text{min}} + T} S_{\text{axion}}^2 (\omega, t) dt d\omega.
\end{align}

The natural line-width of an axion-photon conversion signal in the rest frame of a non-relativistic axion star is $\Delta \omega \sim m_a v^2/(2\pi)$ where $v$ is the axion velocity dispersion. For relativistic axion stars, such as oscillons, the axion spectrum is no longer dominated by a single harmonic. There are peaks at approximately $\omega_0, 3\omega_0, 5\omega_0, \dots$ where $\omega_0$ is the frequency of the fundamental mode, set by the axion mass minus the binding energy \cite{baum2017, marios2019}. Based on the simulations in \cite{marios2019}, the line-width of the fundamental mode is $\Delta \omega = \mathcal{O}(1) m_a/(2\pi)$. There are several broadening effects that can significantly impact the signal width. For axions that are not normally incident upon the critical surface, the rotation of the pulsar can impart momentum to the converted photons, leading to a Doppler-broadened signal \cite{Battye:2019aco}.\footnote{For mis-aligned ($\theta_m \ne 0$) pulsars, Doppler broadening occurs even if axions are normally incident \cite{Battye:2019aco}.} A related effect is the refraction of converted photons propagating in the spatially and temporally varying magnetosphere plasma. This induces a fractional line-width of $\Delta \omega/\omega \sim v_{\text{cs}}$, where $v_{\text{cs}}$ is the velocity of the critical surface \cite{Foster:2020pgt}. For non-relativistic axion stars, these effects can greatly broaden the signal, while for relativistic axion stars, converting well within the light cylinder, they are dominated by the natural line-width of the signal. For high axion masses, the signal width will be larger than the bandwidth available to the radio telescope, which will suppress the signal. The effective bandwidth is given by $\mathcal{B}_{\text{eff}} = \min(\Delta \omega, \mathcal{B}_{\text{rec}})$ where $\Delta \omega$ is the signal bandwidth and $\mathcal{B}_{\text{rec}}$ is the receiver bandwidth. Additionally, since the oscillon is of finite extent, a single event will have duration given by the crossing time of the oscillon across the conversion region $T_{\text{dur}} = L/v_c$ (\ref{eqn:eventduration}) and $v_c$ is the velocity of the oscillon when it reaches the conversion radius. For an observation time $T \gg \tau_{\text{eff}}$, the effective integration time is  $T_{\text{eff}} = \Gamma T \tau_{\text{eff}}$, where $\Gamma$ is the expected event rate occurring in the field-of-view. As mentioned in Sec. \ref{sec:rates}, we are observing the Milky Way bulge, giving an event rate of
 
\begin{align}
    \Gamma &= 2\pi \displaystyle\int_{-\infty}^\infty \displaystyle\int_{0}^{D \sqrt{\text{FoV}}} {\rho_{\text{DM}}\left(\sqrt{s^2 + z^2} \right) \over M_{\text{osc}} } n_{\text{bulge}}\left(\sqrt{s^2 + z^2} \right) \left \langle \sigma v \right \rangle s \ ds dz \label{eqn:rateFoV}
\end{align}

\noindent where $D \approx 8.5$ kpc is the distance to the bulge and FoV is the field-of-view of the relevant telescope. The other parameters are defined in Sec. \ref{sec:rates}. For simplicity, we assume all events have the same flux and spectrum. A more careful analysis would require considering the distribution of pulsar properties and event fluxes when calculating the $\chi^2$ value but this is beyond the scope of this work. 

\begin{figure}
    \centering
    \includegraphics[scale=0.45]{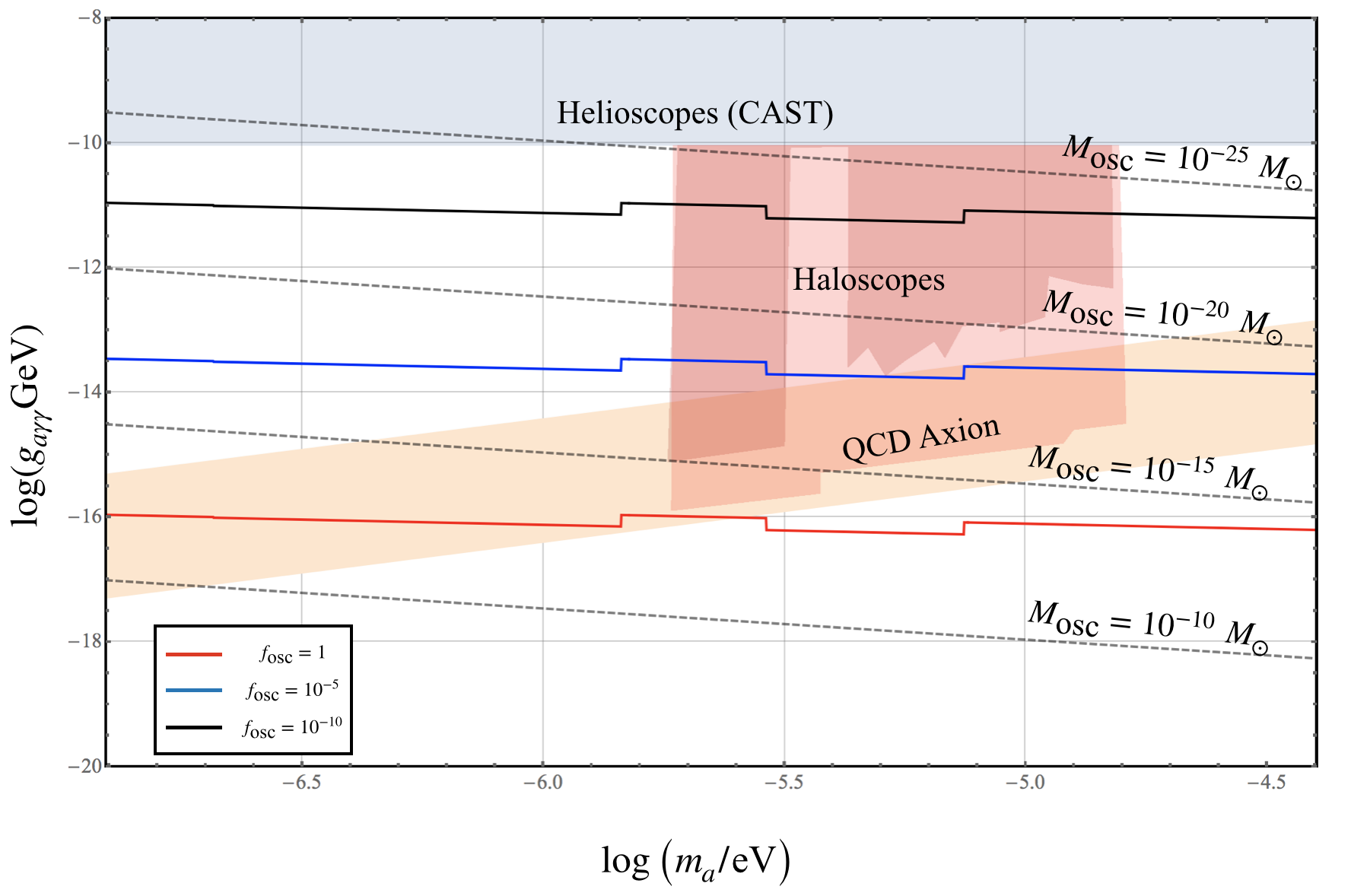}
    \label{fig:Sensitivity1}
    \caption{Sensitivity of various radio missions to signals from ONS conversions. The plots represent the lowest axion-photon coupling that can be detected when the fraction of DM in the form of oscillons is $f_\text{osc} = 1$ (red), $f_\text{osc} = 10^{-5}$ (blue), and $f_{\text{osc}} = 10^{-10}$  (black). At high axion mass the sensitivity is cut off by the requirement that the conversion radius be greater than the neutron star radius. The sensitivity was computed by assuming a one-year observation of the center of the galactic bulge. The low-frequency range ($30$ MHz-$350$ MHz) is covered by LOFAR \cite{LOFAR} and SKA1-LOW, the medium-frequency range ($350$ MHz-$14$ GHz) by SKAMID, and the high-frequency range (14 GHz-50 GHz) by JVLA \cite{SKABaselineDesign}. Projections were made assuming $\gagg = \alpha/(2\pi f_a)$, pulsars with $B = 10^{13}$ Gauss, and rotational period $P = 1$ sec. Oscillon mass contours are also presented for reference (dashed).} 
\end{figure}

We find that, for most radio missions of interest, events that occur within the observing time will be far enough above the detection threshold that they may be detected with high signal-to-noise. Therefore, the limiting condition is that there must be at least one event within the observing window. To optimize for event rate, we consider a combination of radio missions that have the highest FoVs in their respective radio bands. To estimate the sensitivity of various radio observatories, we use the telescope parameters listed \cite{SKABaselineDesign}. The lowest accessible frequency is 30 MHz, which is observed by the LOFAR array \cite{LOFAR}. Between 50 MHz and 14 GHz, SKA1-Low and SKA2-Mid provide continuous coverage as well as high FoV \cite{SKABaselineDesign}. Between frequencies 700 MHz and 1.8 GHz, the ASKAP interferometric array provides a higher FoV than SKA, making it more sensitive to ONS collisions, despite its higher SEFD. High frequencies, between 14 GHz and 50 GHz, can be observed by the Green Bank Telescope (GBT) and the Very Large Array (VLA), the latter proving better suited owing to its considerably larger FoV \cite{SKABaselineDesign}. Based on the discussion above, the reach of various radio missions is set by the the condition $\Gamma = 1\text{ yr}^{-1}$, where $\Gamma$ is given in (\ref{eqn:rateFoV}). This projected sensitivity is plotted in Fig. \ref{fig:Sensitivity1}. We have assumed a neutron star surface magnetic field of $B_0 = 10^{14}$ Gauss, orbital period of $P = 1$ second, and axion-photon coupling $\gagg = \alpha/(2\pi f_a)$. In Fig. \ref{fig:Sensitivity1}, we have included the QCD axion band purely for reference. We emphasize that the models leading to long-lived oscillons are not related to the QCD axion. 

\section{Fast Radio Bursts}
\label{sec:FRB}

In this section we provide some comments on previous proposals that axion star-neutron star collisions could be responsible for the enigmatic Fast Radio Bursts (FRBs). FRBs are very powerful and highly dispersed millisecond pulses of radio emission in the GHz frequency range; the origin of FRBs is yet unknown \cite{Lorimer2007, Keane2012, Thornton2013, Spitler2014, Ravi2015, Petroff2017}. The large dispersion measures  ($>100$ pc cm$^{-3}$) and isotropic distribution away from the galactic plane suggest that the sources may be extragalactic in origin. At the time of writing, there have been approximately 100 detected FRBs, several of which have been shown to repeat. As mentioned in Section \ref{sec:tidal}, there exist numerous papers positing axion star neutron star collisions as the progenitors for FRBs. More recently, there has been a claim that ONS collisions may be the source of FRBs \cite{Buckley:2020fmh}. This is intriguing as oscillons are tidally stable near the conversion radius and an enormous amount of radio energy is released in a collision. However, we argue that this mechanism cannot be responsible for all of the observed FRBs. The first reason is that the time scales of these collisions do not match those of observed FRBs. The time scale is set by the crossing time of the oscillon across the resonant conversion region which is given by \ref{eqn:eventduration}. This gives events that are at least two orders of magnitude shorter in duration than most observed FRBs. The authors of \cite{Buckley:2020fmh} suggest that this timescale can be enhanced if the oscillon travels within the resonant region for $\sim$ milliseconds before inspiraling towards the neutron star. While this scenario is plausible, the event rate will be severely suppressed by phase space considerations, meaning it is unlikely to be able to explain the observed abundance of FRBs.

Another difficulty with this progenitor theory is that FRBs have been observed over a wide range of frequencies spanning $\omega_{\text{FRB}} \in [400 \text{ MHz}, 8 \text{ GHz}]$, while signals from ONS collisions are produced at a small range of frequencies around $m_a$. The authors of \cite{Buckley:2020fmh} argue that the spread of observed frequencies can be explained by a combination of cosmological and gravitational redshift. The former can contribute only a factor of $1+z \lesssim 4$ based on FRB observations \cite{walker2018constraining}. We argue that gravitational redshift does not contribute to the frequency spread of observed signals. This is because the energy of the photons produced at the conversion radius is the energy of the axions at that point. During their infall, the axions accrue kinetic energy that exactly cancels the gravitational redshift term. Therefore, all of the photons produced will have frequency set by the axion mass (minus the binding energy in the oscillon). Thus the observed spread in FRB frequencies cannot be explained by ONS collisions.

\section{Conclusions}

We have revisited the problem of axion stars colliding with pulsar magnetospheres. The inhomogeneous plasma frequency in the magnetosphere gives the photon a space-time dependent mass. When an axion star reaches a critical radius, at which the plasma frequency equals the axion mass, there is a resonant conversion of axions into photons, leading to a dramatic radio signal. Such signatures can be observed by existing radio telescopes. These events, occurring both in the Milky Way as well as in nearby satellite galaxies, can be used to set limits on the abundance of axionic dark matter that exists in the form of oscillons. As future telescopes achieve better sensitivities and higher fields-of-view, less integration time will be necessary to observe such events and hence more stringent constraints can be set on the abundance of oscillons in DM halos. 

Unlike many previous studies, we focus on axion stars that are not tidally disrupted prior to reaching this critical radius. The best candidate that satisfies this condition is the oscillon. We have argued that oscillons arising from models with flat potentials could possibly have very long lifetimes and make up a sizeable fraction of the present-day DM density. We have presented a more detailed calculation of the tidal forces on axion stars in the gravitational fields of pulsars and shown that the disruption radius can be quite a bit smaller than the Roche radius. This leaves open the possibility of the aforementioned effect being observable for some solitons. Future work will include a careful treatment of the tidal forces acting on solitons. 
Even in the event of tidal disruption, the stream that forms around the star can create a longer signal whose maximum power is, however, diminished due to the lower density of axions.

We proposed a scheme to detect ONS events in the galactic bulge, arguing that over a wide range of parameter space, the detection of ONS events is limited by event rate, rather than peak flux density. Our scheme addresses this limitation by emphasizing the importance of field-of-view over intrinsic noise in radio telescopes. The sensitivity of our scheme is presented in Fig. \ref{fig:Sensitivity1}. We find that it is possible to see ONS events even if oscillons make up a very small fraction of dark matter. Finally, we discuss the possibility of this mechanism being responsible for FRBs. While the energetics seem consistent, the timescale is orders of magnitude too small. Additionally, the observed frequency spread of FRBs cannot be explained by ONS collisions at a single axion mass.

\acknowledgments

We are grateful to Asimina Arvanitaki, Sebastian Baum, and Savas Dimopoulos for useful discussions during the preparation of this work, as well as valuable comments on preliminary drafts. This work was supported by the National Science Foundation under Grant No. PHYS- 1720397 and the Gordon and Betty Moore Foundation Grant GBMF7946. AP acknowledges the support of the Fletcher Jones Foundation.

\bibliography{ONS}{}
\bibliographystyle{apsrev4-1}







\end{document}